\documentclass[a4paper,11pt]{article}
\usepackage[top=2.5cm, bottom=2.5cm, left=2.5cm, right=2.5cm]{geometry}
\usepackage{authblk}
\usepackage{hyperref}
\usepackage{tcolorbox}
\usepackage{amsmath}
\usepackage{amssymb}
\usepackage[authoryear,round]{natbib}
\usepackage[utf8]{inputenc}
\usepackage[T1]{fontenc}

\title{Estimating measures of information processing during cognitive tasks using functional magnetic resonance imaging}

\author[1,2,3]{\small Chetan Gohil}
\author[1,3]{\small Oliver M. Cliff}
\author[2,3]{\small James M. Shine}
\author[3,4]{\small Ben D. Fulcher}
\author[1,3]{\small Joseph T. Lizier}

\affil[1]{School of Computer Science, The University of Sydney, New South Wales, Australia}
\affil[2]{Brain and Mind Centre, The University of Sydney, New South Wales, Australia}
\affil[3]{Centre for Complex Systems, The University of Sydney, New South Wales, Australia}
\affil[4]{School of Physics, The University of Sydney, New South Wales, Australia}

\date{}

\begin{document}
\maketitle

\begin{abstract}
Cognition is increasingly framed in terms of information processing, yet most fMRI analyses focus on activation or functional connectivity rather than quantifying how information is stored and transferred. To remedy this problem, we propose a framework for estimating measures of information processing: active information storage (AIS), transfer entropy (TE), and net synergy from task-based fMRI. AIS measures information maintained within a region, TE captures directed information flow, and net synergy contrasts higher-order synergistic to redundant interactions. Crucially, to enable this framework we utilised a recently developed approach for calculating information-theoretic measures: the \textit{cross} mutual information. This approach combines resting-state and task data to address the challenges of limited sample size, non-stationarity and context in task-based fMRI. We applied this framework to the working memory ($N$-back) task from the Human Connectome Project (470 participants). Results show that AIS increases in fronto-parietal regions with working memory load, TE reveals enhanced directed information flows across control pathways, and net synergy indicates a global shift to redundancy. This work establishes a novel methodology for quantifying information processing in task-based fMRI.
\end{abstract}

\section{Introduction}

Cognition is increasingly framed in terms of \textit{information processing}, where the brain can be thought of as a complex dynamical system that encodes, stores, and manipulates information~\citep{neisser2014cognitive}. This perspective underpins major theoretical frameworks such as predictive coding~\citep{rao1999predictive}, the free-energy principle~\citep{friston2010free}, computational models of cognition~\citep{anderson2015cognitive}, and integrated information theory~\citep{tononi2016integrated}. Testing these theories, however, requires quantitative methods for characterising how information is processed in the brain.

In this study, we focus on quantifying different aspects of information processing (storage and transfer) during a cognitive task using functional magnetic resonance imaging (fMRI). FMRI is a non-invasive technique for measuring brain activity. It detects changes in blood oxygenation and flow that are coupled to neural activity, known as the blood-oxygen-level-dependent (BOLD) signal, thereby providing an indirect measure of regional brain function~\citep{logothetis2008we}. Task-based fMRI allows researchers to identify brain regions engaged by specific cognitive processes (e.g. working memory, attention, or decision-making) and to examine how activity patterns change with task demands~\citep{barch2013function}.

Information theory~\citep{cover1999elements} provides a principled framework for quantifying the amount of information within a system of interacting components. Several studies have explored its use in neuroimaging for characterising brain function. Early work calculated the mutual information of fMRI data to characterise inter-regional functional connectivity~\citep{gomez2012analysis, hlinka2011functional}. Later approaches extended this to transfer entropy as a measure of directed connectivity~\citep{lizier2011multivariate, vicente2011transfer, wibral2014directed}, while more recent developments have used partial information decomposition to disentangle redundant and synergistic contributions to neural processing~\citep{wibral2017quantifying, mediano2019beyond, rosas2019quantifying, luppi2022synergistic, varley2023multivariate}. 

The vast majority of previous work has focused on applying information theory to resting-state fMRI data, given that these recordings tend to be longer and are typically considered to better adhere to the assumption of stationarity due to fixed experimental conditions: these characteristics are favourable for estimating the underlying probability distribution needed to calculate information-theoretic measures. Methods for estimating information processing in task-based fMRI are less well developed. Fundamental questions remain about which measures are most appropriate, how to interpret them in the context of cognition, and how best to address the challenges posed by the modality: task-based fMRI recordings are inherently non-stationary~\citep{laumann2024challenges} and typically short. Additionally,  the temporal resolution of fMRI is limited by the slow hemodynamic response function (HRF)~\citep{logothetis2008we}. In this study, we tackle these issues by first deconvolving the HRF~\citep{wu2013blind}, and crucially, in our work we adopt a recently developed approach for handling non-stationarity and short recordings: the \textit{cross} mutual information~\citep{gohil2025cross}.

We will focus on characterising brain activity using three complementary measures of information processing~\citep{lizier2012thelocal, li2019transitions, voges2024decomposing}:
\begin{itemize}
\item Active information storage (AIS), which quantifies how much information about a region’s past is maintained over time~\citep{lizier2012local}.
\item Transfer entropy (TE), which captures directed information flow between regions and can be seen as a non-linear generalisation of Granger causality~\citep{schreiber2000measuring}.
\item Net synergy, which compares how much information arises from the joint contribution of multiple inputs together rather than in isolation.
\end{itemize}
Together, these measures provide a principled way to characterise how information is stored and transferred in the brain during cognitive processes. They also provide the ability to distinguish whether the inter-regional interactions are primarily redundant or synergistic, which further clarifies the role of information processing in cognitive function.

In this work, we will study fMRI data from the Human Connectome Project (HCP), a large-scale resource that provides high-quality fMRI data from healthy individuals~\citep{van2013wu}. We focus on the $N$-back task, a widely used paradigm that engages working memory by requiring participants to monitor a sequence of stimuli and indicate whether the current stimulus matches one presented $N$ steps earlier~\citep{owen2005n}. The $N$-back task robustly activates a fronto-parietal control network (including dorsolateral prefrontal and parietal cortices, the thalamus, and insula) while deactivating the default mode network (DMN), with activity scaling as a function of memory load~\citep{owen2005n, emch2019verbal}. These patterns are reliably observed in the HCP dataset at both the group and individual level~\citep{barch2013function, lamichhane2020exploring}, enabling the large-scale investigation of variability in working memory function. Moreover, the $N$-back task has been used to study dynamic reconfiguration of functional networks under cognitive demand~\citep{shine2016dynamics, dySCo2025}, making it well suited to an information-theoretic analysis: maintaining items in memory relates naturally to AIS, dynamic coordination across fronto-parietal networks reflects TE, and integrating inputs from multiple regions provides a setting to examine synergistic versus redundant interactions.

In Section~\ref{sec:methods}, we describe our methodology for estimating information processing measures from task-based fMRI data, paying particular attention to challenges such as defining an underlying probability distribution for the data. Section~\ref{sec:results} presents results from the $N$-back task in the HCP dataset. Section~\ref{sec:discussion} discusses the implications of our findings, and Section~\ref{sec:conclusions} summarises the contributions and outlook.

\section{Methods} \label{sec:methods}

\subsection{Human Connectome Project (HCP) Dataset} \label{sec:dataset}

In this study, we analyse the working memory ($N$-back) task from the HCP dataset~\citep{barch2013function, van2013wu}. The HCP dataset includes both a 0-back (baseline attention) condition and a 2-back (working memory) condition - a subtraction of 2-back $>$ 0-back enables a direct comparison between high and low cognitive demand~\citep{barch2013function}. In addition, the HCP dataset contains an independent resting-state recording for each subject. We studied the 500 subjects release.

\subsubsection*{Pre-processing}

Bias field correction and motion correction (12 linear degrees of freedom using FSL’s FLIRT) were applied to each subject's fMRI recording (resting-state and $N$-back separately) as part of the minimal pre-processing pipeline~\citep{glasser2013minimal}. Temporal artifacts were identified in each dataset by calculating framewise displacement from the derivatives of the six rigid-body realignment parameters estimated during standard volume realignment~\citep{power2014methods}, as well as the root-mean-square change in BOLD signal from volume to volume (DVARS). Abnormal frames were not excluded from the data.

Following artifact detection, nuisance covariates associated with the six linear head movement parameters (and their temporal derivatives), frame-wise displacement, DVARS, and anatomical masks from the CSF and deep cerebral WM were regressed from the data using the CompCor strategy~\citep{behzadi2007component}. Based on this pipeline, a total of 27 subjects were removed based on significant head motion (framewise displacement $>$ 2.5\% and/or DVARS $>$ 0.25)~\citep{power2014methods}.

\subsubsection*{Parcellation}

Following the pre-processing, a mean time series was extracted from each of the 333 pre-defined regions-of-interest (ROIs): 161 and 162 regions from the left and right hemispheres respectively using the Gordon atlas~\citep{gordon2016generation}.

\subsubsection*{Preparation}

Following the parcellation, we normalised ($z$-scored) and detrended the data. We then applied a zero-phase 3rd order forward-backward Butterworth filter (0.01 $<$ f $<$ 0.08 Hz).  Global signal regression was then applied to the data\footnote{This step did not have a substantial impact on the results presented in this manuscript.}~\citep{murphy2017towards}. To mitigate the effects of the HRF on the information processing measures, we used a blind deconvolution technique to remove the HRF from the data~\citep{wu2013blind}. Two subjects failed the HRF deconvolution and were removed. Additionally, one subject was removed because it did not have a 2-back accuracy score. This resulted in a final total of 470 subjects with resting-state and $N$-back data.

Finally, to avoid issues with edge effects induced by the bandpass filter, we cropped the first 52 and last 55 samples from the $N$-back recording (leaving 298 samples) and the first and last 100 samples from the resting-state recording (leaving 1000 samples). These time series were then used in the subsequent analysis.

\subsection{Information-Theoretic Measures}

Here, we describe the measures used to quantify information processing in this work. We focus on estimating the storage and transfer of information in the brain as measured by fMRI~\citep{bryant2025unifying}. All information-theoretic measures were calculated using the KSG1 estimator~\citep{kraskov2004estimating} in the JIDT toolbox~\citep{lizier2014jidt}. This estimator overcomes limitations of other kernel-based estimators that suffer from binning bias or restrictive assumptions about form of the probability distribution for the data.

\textbf{Autocorrelation}. Information-theoretic measures often assume that the underlying time series are independent and identically distributed (i.i.d.) across samples~\citep{lizier2012local, bossomaier2016transfer}. FMRI data, however, are strongly autocorrelated due to both intrinsic neural dynamics and the slow HRF. This autocorrelation can lead to inflated estimates of information-theoretic measures, as dependencies in the signal may arise trivially from temporal smoothness rather than genuine information processing~\citep{cliff2021assessing}. Preprocessing was applied to the data to mitigate these issues (detrending, filtering, HRF deconvolution; see Section~\ref{sec:dataset}). Furthermore, in calculating the information-theoretic measures we excluded autocorrelated samples~\citep{theiler1986spurious} to reduce biases related to autocorrelation in the estimator (\texttt{DYN\_CORR\_EXCL}=15 in JIDT, selected to be larger than the length over which any of the time-series analysed retains a statistically significant autocorrelation).

\subsubsection*{Mutual Information (MI)}

\textbf{Cross MI}. Each measure is based on the MI,~\citep{cover1999elements, gohil2025cross}
\begin{equation} \label{eq:mi}
I(X; Y) = \mathbb{E}_{x,y \sim p(x,y)} \left\{ i(x,y) \right\} = \mathbb{E}_{x,y \sim p(x,y)} \left\{ \log \left( \frac{q(x,y)}{q(x) q(y)} \right) \right\}
\end{equation}
where $i(x,y) = \log \left( \frac{q(x,y)}{q(x) q(y)} \right)$ is the \textit{local} (or \textit{pointwise}) MI between samples $x$ and $y$ of the random variables $X$ and $Y$ respectively, $q(x,y)$ is the \textit{reference} distribution and $p(x,y)$ is the \textit{test} distribution. Here, the MI is averaged over data samples from the test distribution $p(x,y)$ but the probabilities given by the reference distribution $q(x,y)$ in the calculation of the local MI. Note, by calculate the MI in this way we are adopting a \textit{cross} MI~\citep{gohil2025cross}. This is an important choice to ensure that we correctly account for the possible \textit{states} ($x,y$ values) the system could explore using the reference distribution.

\textbf{Reference distribution}. This distribution is intended to characterise the full probability of states ($x,y$ values) we expect the brain to exhibit. For this, we propose using the empirical distribution calculated from all the fMRI data we have available for a subject (here both resting state and $N$-back task). Including the resting-state recording in this reference distribution provides the proper context for the $N$-back conditions. It allows us to account for the fact that the fMRI data during the $N$-back conditions may occupy a low-probability region of the state space. Moreover, inclusion of all of the available recordings provides enough baseline data for the probability distributions to be reliably estimated.

\textbf{Test distribution}. This is the distribution we take the expectation over in Equation~\eqref{eq:mi}. The cross MI evaluates how strongly the relationship between $X$ and $Y$ (as defined by the reference distribution) is actually expressed in the test distribution. For this, to evaluate the cross MI for a particular task one would use the task data alone to estimate the test distribution. Here, we will separately use each of the 2-back, 0-back and resting-state data.

\textbf{Conditional MI}. The above approach based on the cross MI is in contrast to a conventional MI calculation, which would use the test distribution as the reference, i.e. $q(x,y) = p(x,y)$. In the context of fMRI task analysis, the task data would be used to define an empirical distribution that would be used for $p(x,y)$ and $q(x,y)$. This choice for the reference distribution defines a probability space based on the data recorded during the task only. It is effectively \textit{conditioning} on the task~\citep{gohil2025cross}, which means it does not have the proper context to know whether the $x$-$y$ dependencies in the task condition are atypical (have low probabilities) compared to a baseline resting-state condition. We call this conventional MI calculation the \textit{conditional MI} because it is based on a probability distribution that is conditioned on the task. Additionally, this approach suffers from the reference distribution $q(x,y)$ being determined from very short and non-stationary data.

\subsubsection*{Cross vs Conditional MI}

All measures calculated in this work will be based on the cross MI. Adopting this approach for task-based fMRI analysis is one of the key innovations of this paper. The characteristics/features of the cross MI compared to a conditional MI is studied in detail using simulated data in~\citep{gohil2025cross}. As outlined in~\citep{gohil2025cross}, a large positive local value for the cross MI $i(x,y)$ means that, under the reference distribution, the joint occurrence $(x,y)$ is more probable relative than would be expected by independence; the expectation over task samples (the test distribution) then tells us whether such (potentially rare) co-occurrences happen systematically during the task. Comparing the cross and conditional MI:
\begin{itemize}
\item A high cross MI and conditional MI indicates a strong $X$-$Y$ dependence within task that is also generally observed at rest.
\item A high cross MI but low conditional MI indicates $X$-$Y$ dependencies that are rare at rest occur frequently in the task. This can happen when a small number of highly informative, task-specific events occur. Here, conditioning on the task removes a redundancy which weakens the $X$-$Y$ relationship measured by the conditional MI.
\item A low cross MI but high conditional MI indicates a strong $X$-$Y$ dependency in task that is not generally observed at rest. Here, conditioning on the task reveals a synergy that strengthens the $X$-$Y$ relationship measured by the conditional MI.
\end{itemize}
In the analysis of task-based fMRI, responses are context-dependent: a dependence that looks strong when only looking at task data might still be part of the brain's normal activity that would be observed at rest. The cross MI lets us ask whether task events (in the test distribution) represent atypical $X$-$Y$ dependencies relative to the subject's broader baseline activity (which is characterised by the reference distribution). The cross MI is a useful measure to have when asking ``does the task induce connectivity at this is unusual compared with baseline?''.
 
To calculate the cross MI in JIDT, we provide both the $N$-back task and resting-state data for the reference distribution (in this work implemented by temporally concatenating the $N$-back task and resting-state data). Then, the local values $i(x,y)$ are estimated to give a time-varying estimate of the information-theoretic measure. In cases where we calculate the conditional MI instead, we will state this explicitly.

\subsubsection*{Active Information Storage (AIS)}

We propose using the AIS as a measure of information storage~\citep{lizier2012local},  
\begin{equation} \label{eq:ais}
A_X = I(X_t; X_{<t}),
\end{equation}
where $X$ is the time series data for a particular ROI. The AIS quantifies the extent to which the current time point $X_t$ can be predicted by the past $X_{<t} = \{ X_{t-1}, X_{t-2}, ... \}$, with higher values indicating that more information is being actively maintained over time. Unlike autocorrelation-based measures, the AIS provides an information-theoretic quantification of temporal dependence that is sensitive to both linear and non-linear dynamics, and accumulates all information provided by multiple past time points together rather than individually. As such, AIS has been widely deployed in studies of information storage in neural data~\citep{wibral2014local, voges2024decomposing, brodski2017information, puche2020active}.

To calculate the AIS, we need to specify the number of previous time points in $X_{<t}$ to use in the MI calculation. In this work, we used 2 time steps (\texttt{k\_HISTORY}=2 in JIDT) which provides the maximum bias-corrected value for the time-series here~\citep{garland2016leveraging, erten2017criticality}.

\subsubsection*{Transfer Entropy (TE)}

To quantify information flows, we propose using the TE~\citep{bossomaier2016transfer, wibral2014directed},  
\begin{equation} \label{eq:te}
T_{X \rightarrow Y} = I(Y_t; X_{<t} | Y_{<t}),
\end{equation}
where $X$ and $Y$ are the time series for a \textit{source} and \textit{target} ROI respectively.  
The TE is a non-linear generalisation of Granger causality~\citep{schreiber2000measuring}, capable of capturing directed, time-lagged, and potentially non-linear dependencies between brain regions. Unlike traditional correlation-based measures (e.g. Pearson correlation), TE accounts for the history of both the source and the target, thus reflects how much new information the source $X_{<t}$ contributes to predicting the future of the target $Y_t$ in the context of what the target's own past $Y_{<t}$ already provides. This makes TE particularly well-suited for studying directed interactions in fMRI and other neural data~\citep{lizier2011multivariate, vicente2011transfer, wibral2014directed, marinazzo2012information, voges2024decomposing}.

To calculate the TE, we need to specify the number of time points to use for the history of the source and the target. In this work we used 2 time steps for both (\texttt{k\_HISTORY}=2 and \texttt{l\_HISTORY}=2 in JIDT) based on first optimising the target past as per AIS and then similarly identifying the source past (specifically the most common value across pairs) to maximise the bias-corrected TE.

\subsubsection*{Net Synergy}

Net synergy (also known as \textit{interaction information})~\citep{mcgill1954multivariate, williams2010nonnegative} provides a way of comparing how synergistic information from two sources contribute to predicting the future of a target (above and beyond the information attributable to one or other source alone) against the redundant information (that may be found in either source alone)~\citep{mcgill1954multivariate}. Here we consider the net synergy between the target's own past as one source itself (being the source for the AIS to the target), and a separate ROI as the other source (being the source of the TE to the target), similar to other studies~\citep{williams2011generalized, wibral2017quantifying, lizier2013towards}. Our measure for net synergy for these sources is
\begin{equation} \label{eq:net-synergy}
S_{X \rightarrow Y} = T_{X \rightarrow Y} - I(X_{<t} ; Y_t) = I(Y_t; X_{<t} | Y_{<t}) - I(X_{<t}; Y_t),
\end{equation}
where $X$ and $Y$ are the source and target time series for the TE respectively. Crucially, this measure is a net between \textit{synergistic} information, i.e. information that is only revealed when the source and target histories are considered together, and \textit{redundancy}, i.e. overlapping information present in both the source and the target's past. A positive value for $S_{X\rightarrow Y}$ indicates that more synergy is provided than redundant information (and indeed is a sufficient condition to conclude that synergistic information is provided), whereas a negative value indicates that more redundant than synergistic information is provided (and is thus a sufficient condition to conclude that redundant information is provided). This measure thus allows us to move beyond simply quantifying directed interactions, and towards understanding whether information exchange between regions is synergetic or redundant~\citep{williams2010nonnegative, rosas2019quantifying, mediano2019beyond}.

To calculate the net synergy, we need to specify the number of time points to use for the history of the source and the target. In this work, we used 2 time steps for both (\texttt{k\_HISTORY}=2 and \texttt{l\_HISTORY}=2 in JIDT) in parallel with the estimates of the AIS and TE.

\subsection{Task Analysis}

\begin{figure}[!ht]
    \centering
    \includegraphics[width=0.8\textwidth]{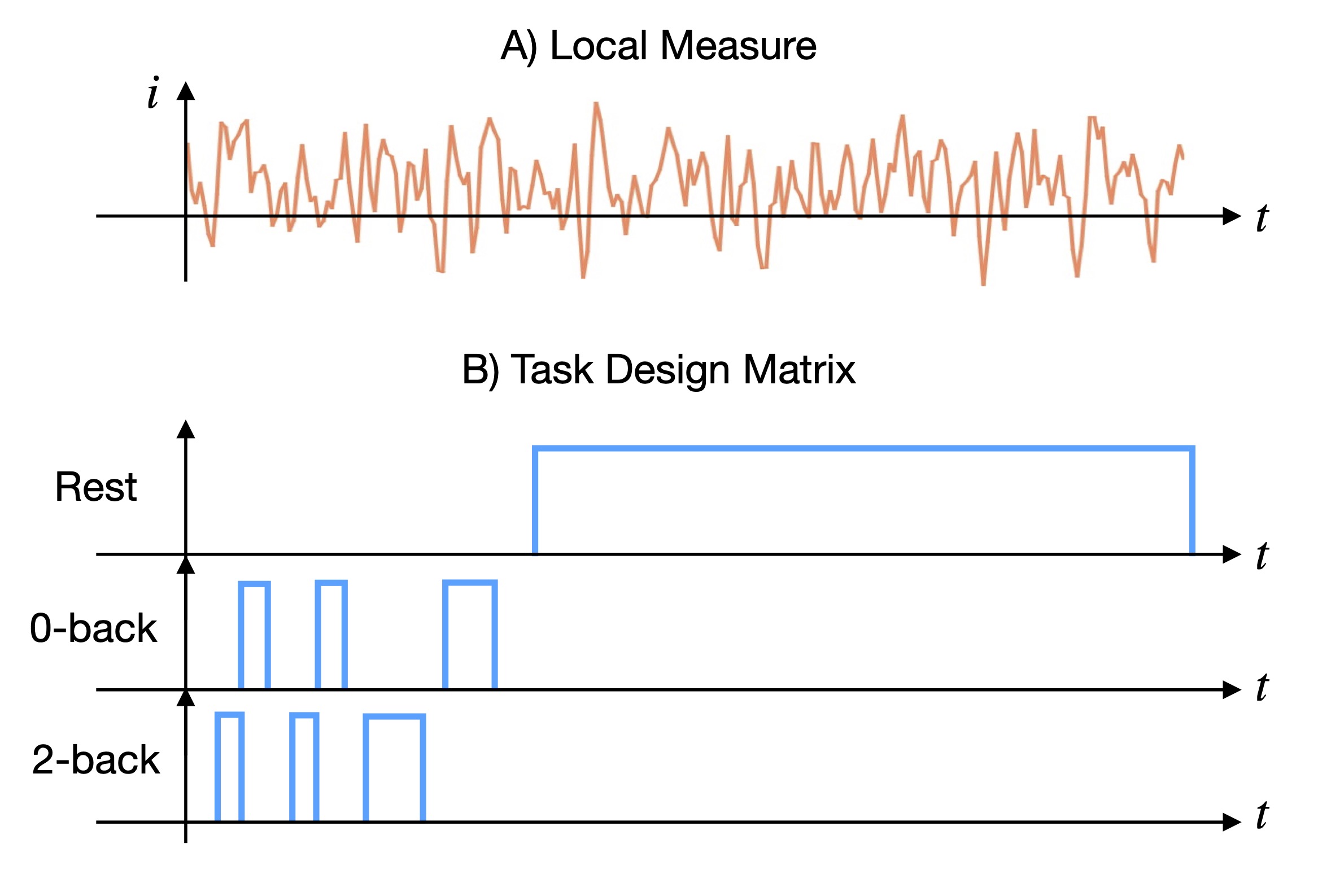}
    \caption{\textbf{Calculation of condition-specific measures}. A) The time series for some local (instantaneous) measure i. Examples include the BOLD signal, local AIS (for a region), and local MI (for an edge). B) Visualisation of the task design matrix where the blue line indicates the time points corresponding to the rest, 0-back and 2-back condition.}
    \label{fig:method}
\end{figure}

Our objective is to estimate $A_X$, $T_{X \rightarrow Y}$, and $S_{X \rightarrow Y}$ for each $N$-back condition (2-back, 0-back) as well as the resting-state condition. To do this, we use a similar approach to a conventional task mean activation analysis~\citep{lindquist2008statistical}. First, we estimate a local (instantaneous) measure at each time point (Figure~\ref{fig:method}A). Then, we fit a first-level General Linear Model (GLM), which contains a regressor for each condition of interest (Figure~\ref{fig:method}B). Note, because we deconvolved our fMRI data we do not need to convolve our task design matrix with the HRF. Instead, we have block (boxcar) regressors in our design matrix. This means we can calculate condition-specific measures by simply taking the average value of the local measure over the time points corresponding to each condition. Each region/edge has an independent first-level GLM.

\subsubsection*{First-Level Contrasts}

After calculating condition-specific measures for each subject (2-back, 0-back, rest), we computed two first-level contrasts:
\begin{enumerate}
\item 2-back vs rest (2-back minus rest). This captures each subject’s response to increased working memory load relative to a resting baseline, highlighting regions and interactions engaged by the task.
\item 2-back vs 0-back (2-back minus 0-back). This accounts for sensory, motor, and attentional processes present in both conditions, isolating neural and information-processing changes specific to the working memory demands of the $N$-back task.
\end{enumerate}

\subsection{Statistical Significance Testing}

\subsubsection*{Group Mean}

To assess statistical significance at the group level, we used non-parametric permutation testing within a GLM framework~\citep{winkler2014permutation}. A group-level GLM with a single mean regressor was fitted to the first-level contrasts, and sign-flip permutations were used to generate the null distribution under the hypothesis that the group mean equals zero. To control for multiple comparisons we used a maximum $t$-statistic correction (over regions/edges).

\subsubsection*{Task Performance (2-Back Accuracy)}

To examine whether inter-individual variability in task performance (2-back accuracy) was associated with the information processing measures, we fitted a group-level GLM with ($z$-scored) 2-back accuracy scores as a regressor as well as a mean regressor. We used sign-flip permutations to generate a null distribution under the hypothesis of no relationship between an information processing measure and task performance. To control for multiple comparisons we used a maximum $t$-statistic correction (over regions/edges).

\section{Results} \label{sec:results}

\subsection{Conventional mean activation and FC analyses localise changes but do not characterise information processing}

\begin{figure}[hbt!]
    \centering
    \includegraphics[width=\textwidth]{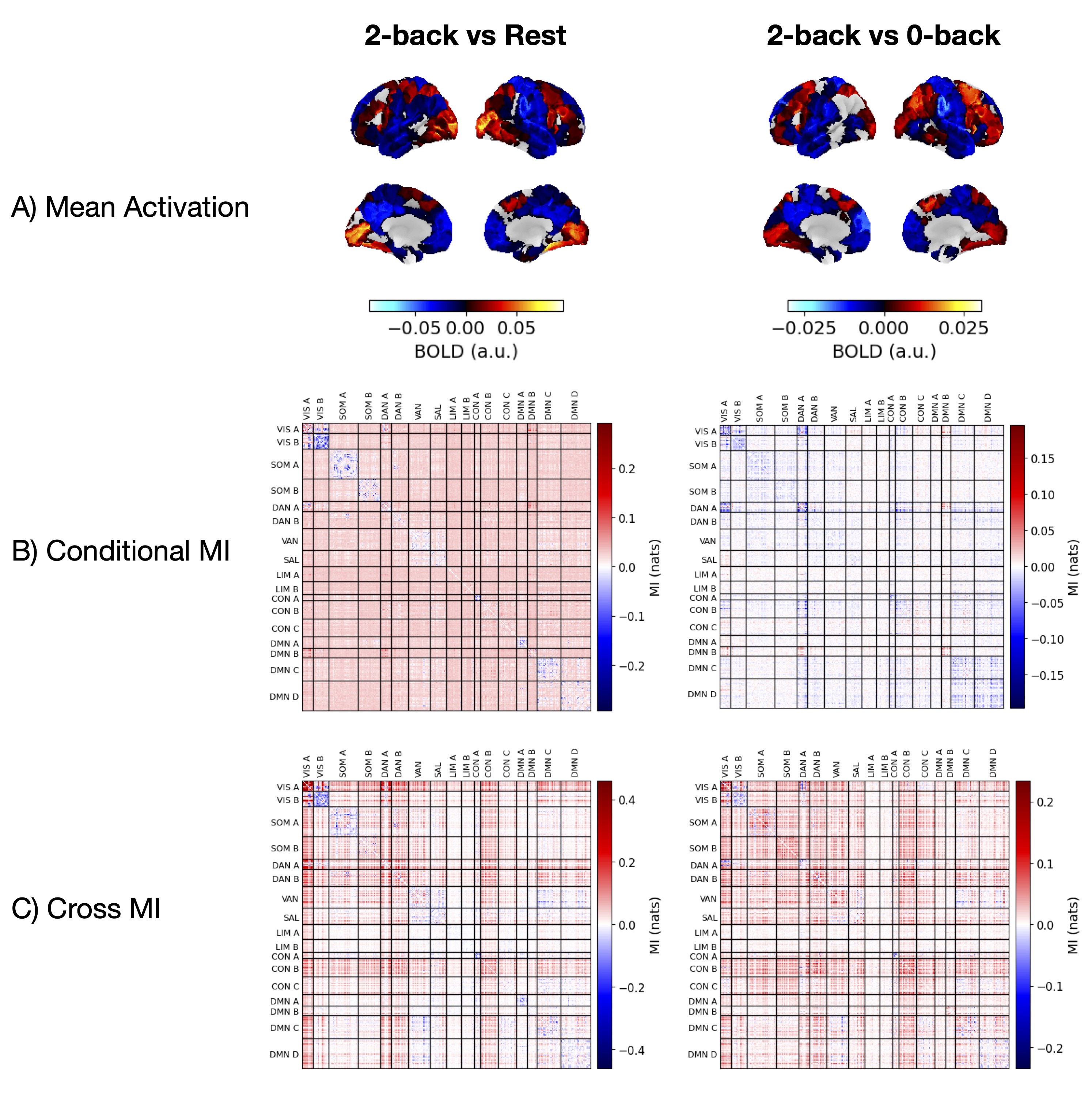}
    \caption{\textbf{Conventional measures for studying task fMRI data}. For the 2-back vs rest condition (left) and 2-back vs 0-back condition (right): A) Change in mean activity, averaging over subjects. B) Change in conditional MI, averaged over subjects. This was calculated for using a conventional approach with the task data defining the reference distribution. C) Change in cross MI, averaged over subjects. This was calculated using the new approach using the resting-state and task data for the reference distribution. All plots have been thresholded to only show contrasts with $p$-values$\,<0.05$. $P$-values were calculated using group-level GLM sign-flip permutations, taking the maximum $t$-statistic over regions/edges to control for multiple comparisons.}
    \label{fig:conventional-measures}
\end{figure}

First, we examine the task response using conventional measures. Figure~\ref{fig:conventional-measures} shows the change in BOLD signal (mean activation) and functional connectivity (FC) using MI for the 2-back vs rest and 2-back vs 0-back contrast. 

The mean activation shows the expected spatial pattern for changes from previous studies~\citep{barch2013function}. In the 2-back condition, we see: the activity in the visual region increases; the activity in fronto-parietal regions increases; and the activity in default mode regions decreases. Note, a key difference between our study and previous work is that we have deconvolved the HRF from the fMRI data. Figure~\ref{fig:conventional-measures}A suggests this step has not significantly altered the spatial activity patterns invoked by the task.

For our measure of FC, we use the conditional and cross MI, which are shown in Figures~\ref{fig:conventional-measures}B and C respectively. We see that the choice of the reference distribution in the MI calculation has had a substantial impact on the change in FC for each contrast (in alignment with numerical experiments presented in~\citep{gohil2025cross}). Looking at the cross MI (Figure~\ref{fig:conventional-measures}C), we see larger FC changes for the 2-back vs rest contrast compared to 2-back vs 0-back and a consistent direction for FC changes across both contrasts. In comparison, the conditional MI (Figure~\ref{fig:conventional-measures}B) shows a similar magnitude for changes in FC for both contrasts whilst the direction of FC changes differs (2-back vs rest generally increases whereas 2-back vs 0-back shows decreases in FC). Since we would expect consistent changes due to working memory in both contrasts, these results suggest that including the resting-state data in the reference distribution (i.e. using the cross MI) provides a better context to observe changes in the 2-back condition compared to rest and the 0-back condition.

The conventional measures shown in Figure~\ref{fig:conventional-measures} indicate substantial changes in FC for the regions involved in the $N$-back task. However, they do not provide any insight into the nature of the activity changes, i.e. what changes in information processing occur at each region. In the subsequent analysis, we use information-theoretic measures to explore how information storage and transfer change in the $N$-back task.

\newpage
\subsection{Information processing changes are organised by modules}

\begin{figure}[hbt!]
    \centering
    \includegraphics[width=\textwidth]{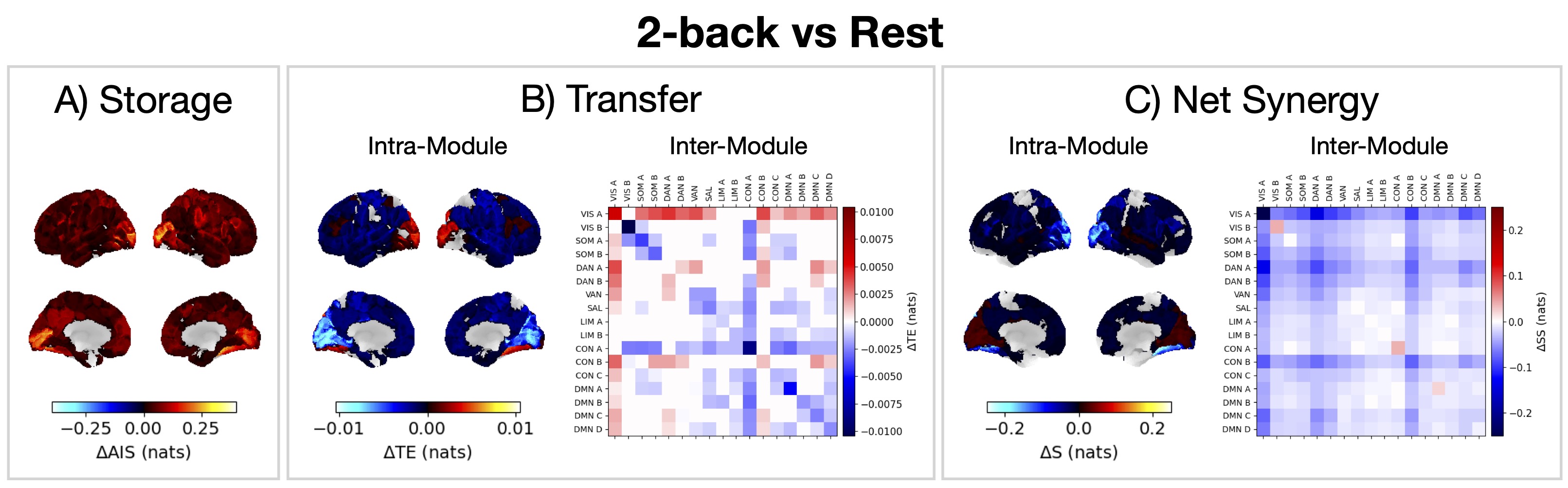}
    \caption{\textbf{Information-theoretic analysis reveals modular reorganisation during the 2-back task (2-back vs rest).} A) Change in AIS (calculated using Equation~\eqref{eq:ais}) averaged over subjects. We see that AIS increases globally, especially in visual modules. B) Change in TE (calculated using Equation~\eqref{eq:te}) averaged over subjects and edges that correspond to Yeo modules. We see that TE decreases across most modules but increases in VIS A, DAN A and CON B. C) Change in net synergy (calculated using Equation~\eqref{eq:net-synergy}) averaged over subjects and edges that correspond to Yeo modules. We see that net synergy shifts towards redundancy across modules, except for VIS B and CON A. Only contrasts with $p<0.05$ are shown. $P$-values were calculated using group-level GLM sign-flip permutations, taking the maximum $t$-statistic over regions/edges to control for multiple comparisons.}
    \label{fig:2-back-vs-rest}
\end{figure}

Figure~\ref{fig:2-back-vs-rest} shows the AIS, TE and net synergy for the 2-back vs rest contrast. Full parcel-by-parcel matrices for the TE and net synergy are shown in Figure~\ref{fig:all-te} and Figure~\ref{fig:all-synergy}. Each measure shows changes that are roughly homogenous across a functional module as defined by the Yeo atlas~\citep{yeo2011organization}. Therefore, for the TE and net synergy we average over the edges assigned to a Yeo module in Figures~\ref{fig:2-back-vs-rest}B and C to obtain a module-level response for each contrast. 

Information storage (AIS; Figure~\ref{fig:2-back-vs-rest}A) increases globally with the largest increases in the visual modules (VIS A and B). Information transfer (TE; Figure~\ref{fig:2-back-vs-rest}B) shows inter-module transfer (in and out) all decrease (in particular those involving CON A) with the exception of the visual module (VIS A), the attention module (DAN A), and the control module (CON B), which show increases. For transfers within each module, all regions except VIS A decrease, most substantially VIS B, CON A and DMN A. This suggests a dominant outflow of information from the visual region to all other regions due to the (visual) stimuli in the $N$-back task.

Figure~\ref{fig:2-back-vs-rest}C shows the change in net synergy for each module. Negative values indicate there is a global shift to redundancy for almost all modules, apart from VIS B, CON A, and DMN A, where there is a shift towards synergistic information transfers within these modules. The synergistic transfers within VIS B, CON A and DMN A are particularly interesting, since it contrasts with the pairwise TE within these regions (summing unique information and redundant information with the target past) which decreases.

The 2-back vs rest contrast includes functional changes associated with the visual input and motor response of the $N$-back task. The changes in information processing due to the visual stimulus dominates in this contrast. In the next section we isolate changes associated with working memory by looking at the 2-back vs 0-back contrast.

\subsection{Working memory increases storage and transfer with a shift to redundancy}  

Here, the baseline condition (0-back) controls for visual stimuli, motor responses, and attention, allowing the 2-back vs 0-back contrast to isolate information-processing changes related to maintaining an item in working memory.

\begin{figure}[hbt!]
    \centering
    \includegraphics[width=\textwidth]{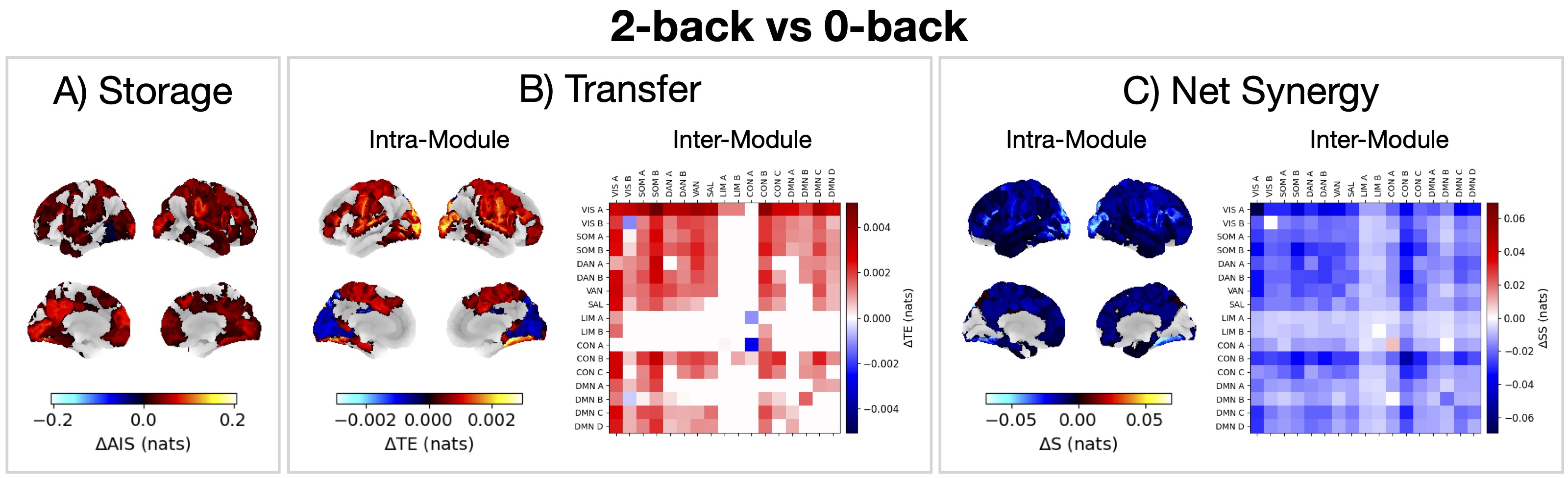}
    \caption{\textbf{Working memory enhances storage and transfer but induces a global redundancy shift (2-back vs 0-back).} A) Change in AIS (calculated using Equation~\eqref{eq:ais}) averaged over subjects.. We see that AIS increases globally. B) Change in TE (calculated using Equation~\eqref{eq:te}) averaged over subjects and edges that correspond to Yeo modules. We see that TE increases broadly across modules except notably VIS B and CON A, which show decreases. C) Change in net synergy (calculated using Equation~\eqref{eq:net-synergy}) averaged over subjects and edges that correspond to Yeo modules. We see that net synergy shifts to redundancy except notably within CON A, which shows increased synergy. Only contrasts with $p<0.05$ are shown. $P$-values were calculated using group-level GLM sign-flip permutations, taking the maximum $t$-statistic over regions/edges to control for multiple comparisons.}
    \label{fig:2-back-vs-0-back}
\end{figure}

Information storage increases in fronto-parietal regions (AIS; Figure~\ref{fig:2-back-vs-0-back}A), mirroring the mean activity changes found in Figure~\ref{fig:conventional-measures}A. Increased AIS indicates greater temporal stability of activity in regions engaged by working memory demands, consistent with the view that working memory depends on the maintenance of neural representations over time~\citep{constantinidis2016neuroscience}.

Information transfers increase broadly across the cortex (TE; Figure~\ref{fig:2-back-vs-0-back}B). Increased information transfer between modules suggests that working memory is a distributed process that involves inter-module communication. Notably, within-module information transfer is reduced within the VIS B and CON A modules, indicating that working memory demands selectively reduce internal communication within these modules. This pattern may reflect a reorganisation toward more local processing, with information being maintained within sub-regions of these modules with less propagation internally. Interestingly, the net synergy shown in Figure~\ref{fig:2-back-vs-0-back}C demonstrates that the smaller within module information transfers in CON A actually skew more strongly towards synergistic combinations or higher-order components, whereas the rest of the cortex broadly shifts to redundant information transfers.

\subsection{Individual performance is related to within-module information storage and a shift to redundancy}

\begin{figure}[hbt!]
    \centering
    \includegraphics[width=0.55\textwidth]{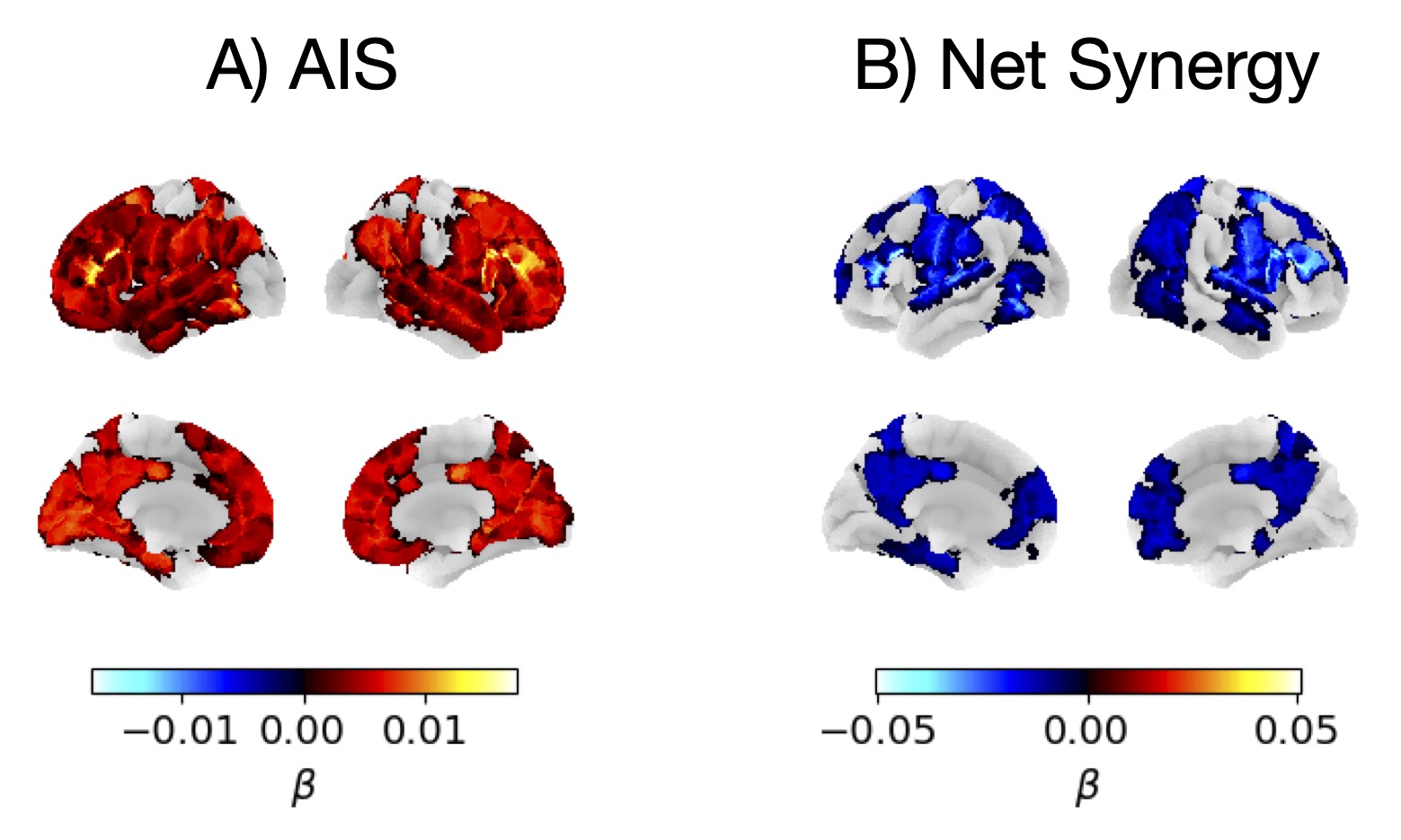}
    \caption{\textbf{Within-module information processing changes predicts task performance.} Regression coefficients ($\beta$) for predicting response accuracy with: A) Change in AIS (2-back vs rest) for each subject. We see greater change in AIS, particularly in frontal regions is linked to higher accuracy. B) Change in net synergy (2-back vs rest) for each subject. We see a stronger shift towards redundancy is linked to better performance. Only coefficients with $p<0.05$ are shown. $P$-values were calculated using group-level GLM sign-flip permutations, taking the maximum $t$-statistic over regions/edges to control for multiple comparisons.}
    \label{fig:2-back-accuracy}
\end{figure}

To examine the behavioural relevance of information processing changes, we regressed within-module changes in information-theoretic measures for the 2-back vs 0-back contrast against individual 2-back accuracy (Figure~\ref{fig:2-back-accuracy}). Higher 2-back accuracy was associated with a greater increase in AIS, an effect that was present globally but strongest in frontal modules. In addition, better performance was associated with a stronger shift towards redundancy, reflected by more negative net synergy values. No significant associations were observed for TE.

Working memory requires task-relevant neural representations to be maintained over time in the presence of ongoing sensory input and interference, a function commonly attributed to activity in prefrontal regions~\citep{miller2001integrative, constantinidis2016neuroscience}. Here, we observe higher AIS, particularly in prefrontal regions, is associated with higher 2-back accuracy, supporting this view. Higher 2-back accuracy was also associated with a shift towards redundancy, indicating a reorganisation of information processing into more overlapping and distributed representations. Redundancy reflects the sharing of information across multiple regions and has been proposed as a mechanism for supporting robustness to noise and perturbations in neural systems \citep{rosas2019quantifying, mediano2019beyond}. In contrast, TE did not significantly predict 2-back accuracy, suggesting that individual differences in working memory performance are more closely related to how information is maintained rather than transferred. Taken together, these findings suggest that working memory performance is associated with the maintenance of neural representations and a shift towards robust, redundant representations.

\section{Discussion} \label{sec:discussion}

In this study, we applied information-theoretic measures to task-based fMRI data from the HCP dataset to quantify information storage, transfer, and net synergy during the $N$-back working memory task. We observed changes in information processing are organised at the level of large-scale functional modules, with visual and control modules playing important roles. Working memory demands lead to global increases in information storage and transfer, which are accompanied by a shift towards redundancy. Individual 2-back task performance can be predicted by increased storage in frontal regions and a shift to redundancy. In all, these findings establish the feasibility and utility of information-theoretic approaches for studying cognitive processes with task-based fMRI.

\subsection*{Methodological Contributions}

The application of information theory to fMRI poses specific challenges: autocorrelation and inter-regional time lags due to the HRF, as well as the limited temporal resolution~\citep{logothetis2008we}. To address issues related to the HRF, we deconvolved the HRF using the approach from~\citep{wu2013blind} before estimating information-theoretic measures. Our results show that this step preserved conventional activation patterns (Figure~\ref{fig:conventional-measures}), while enabling more interpretable estimates of information dynamics.

Another key challenge is the estimation of the appropriate underlying probability distribution of the data to calculate the information-theoretic measures. We resolved this by adopting the novel cross MI approach introduced in~\citep{gohil2025cross}. This approach separates two questions that the conventional, conditional MI conflates: (i) how strongly $X$ and $Y$ depend within the task and (ii) whether this dependency is atypical relative to the subject's broader baseline activity (i.e. rest). It is also able to handle short and non-stationary task data via its use of the longer resting-state data recording in the reference distribution to provide a more wholistic view of baseline brain activity. The cross MI is particularly useful for studying whether a task evokes an $X$-$Y$ dependency that is atypical for a given baseline. Note, we do not propose the cross MI as a universal replacement for conditional MI, the two measures answer different empirical questions as per ~\citep{gohil2025cross}. Throughout this manuscript we treat cross MI as the primary measure because our goal is to identify task-evoked $X$-$Y$ dependencies relative to a baseline (in our case based on the combined resting-state and task data for a subject).

When adopting the cross MI approach, the choice of reference distribution is important~\citep{gohil2025cross}. We compared cross MI (including resting-state data in the reference) and conditional MI (using task-only data for the reference) in Figure~\ref{fig:conventional-measures}. The cross MI produced more consistent and interpretable contrasts, suggesting that including resting-state data in the baseline provides essential context for identifying task-evoked changes. This highlights the importance of carefully defining baselines in task-based analyses, echoing broader concerns in the fMRI literature~\citep{stark2001zero}.

In sum, cross MI provides a principled framework to place task-evoked $X$-$Y$ dependencies into the context of a particular baseline (in our case estimated using resting-state and task combined), allowing the identification of task-specific or rare $X$-$Y$ dependencies that the conditional MI may miss.

\subsection*{Information Processing Related To Working Memory}

Our results provide new insights into the changes in information processing related to working memory. Conventional analyses have consistently implicated a fronto-parietal control network in $N$-back performance, with additional involvement of visual, thalamic, and insular regions~\citep{owen2005n, emch2019verbal}. By moving beyond activation and FC, we show that working memory is characterised by widespread increases in information storage (AIS) and transfer (TE). These effects suggest that maintaining items in working memory requires both stronger temporal stability of representations within regions and enhanced communication between regions.  

Another finding was the global shift towards redundancy in net synergy during 2-back vs 0-back contrast. Redundancy implies overlapping information across inputs, which may contribute to robustness in neural coding and the maintenance of task-relevant representations under high load. At the same time, specific modules (VIS B, CON A) showed increases in synergy, suggesting that localised integrative processing is preserved or enhanced where flexible combination of inputs is required. This aligns with theories of working memory as a dynamic interplay between stable maintenance and flexible updating~\citep{d2002neural, constantinidis2016neuroscience}.

Individual differences in information processing were linked to task performance (Figure~\ref{fig:2-back-accuracy}). Participants with greater changes in AIS in frontal modules performed better in the 2-back task, consistent with the central role of prefrontal cortex in supporting working memory~\citep{miller2001integrative}. In addition, better performers showed a stronger shifts to redundancy. Together, these results suggest that information-theoretic measures can provide behaviourally meaningful markers of cognitive function beyond conventional activation or FC analyses.

\subsection*{Limitations and Future Directions}

Several limitations of this work should be acknowledged: the temporal resolution of fMRI remains a fundamental constraint in studying information dynamics; the estimation of higher-order information-theoretic measures, such as synergy and other information decomposition techniques, remains computationally challenging; and our focus on module-level averaging may obscure fine-grained interactions. Moreover, our analysis remains focussed on the net of synergy and redundancy, since direct measures for each are contested~\citep{lizier2018information}.

While the $N$-back task provides a robust probe of working memory, future work should assess changes in information processing in other cognitive domains, such as attention or decision-making. Future directions also include applying this methodology for assessing information processing to electrophysiological data, e.g. magnetoencephalography, which is a near-direct measure of neuronal activity and has a much higher temporal resolution~\citep{baillet2017magnetoencephalography}.

\section{Conclusions} \label{sec:conclusions}

We demonstrate that information-theoretic measures can be meaningfully applied to task-based fMRI, revealing modular changes in information storage, transfer, and net synergy during working memory. Our findings suggest that working memory enhances storage and transfer globally, while reorganising interactions towards redundancy, with selective increases in synergy in control and visual modules. Furthermore, better individual task performance is associated with increased changes in information storage and a stronger shift to redundancy. These results provide new insight into the changes in information processing related to working memory and establish a novel framework for applying information theory to the study of cognition in task-based fMRI.

\bibliographystyle{apalike}
\bibliography{references}

\clearpage

\setcounter{figure}{0}
\counterwithin{figure}{section}
\counterwithin{table}{section}

\renewcommand\thefigure{S\arabic{figure}}
\renewcommand{\thetable}{S\arabic{table}}

\setcounter{equation}{0}

\section*{Supplementary Information (SI)}
\pagestyle{empty}

\begin{figure}[hbt!]
    \centering
    \includegraphics[width=\textwidth]{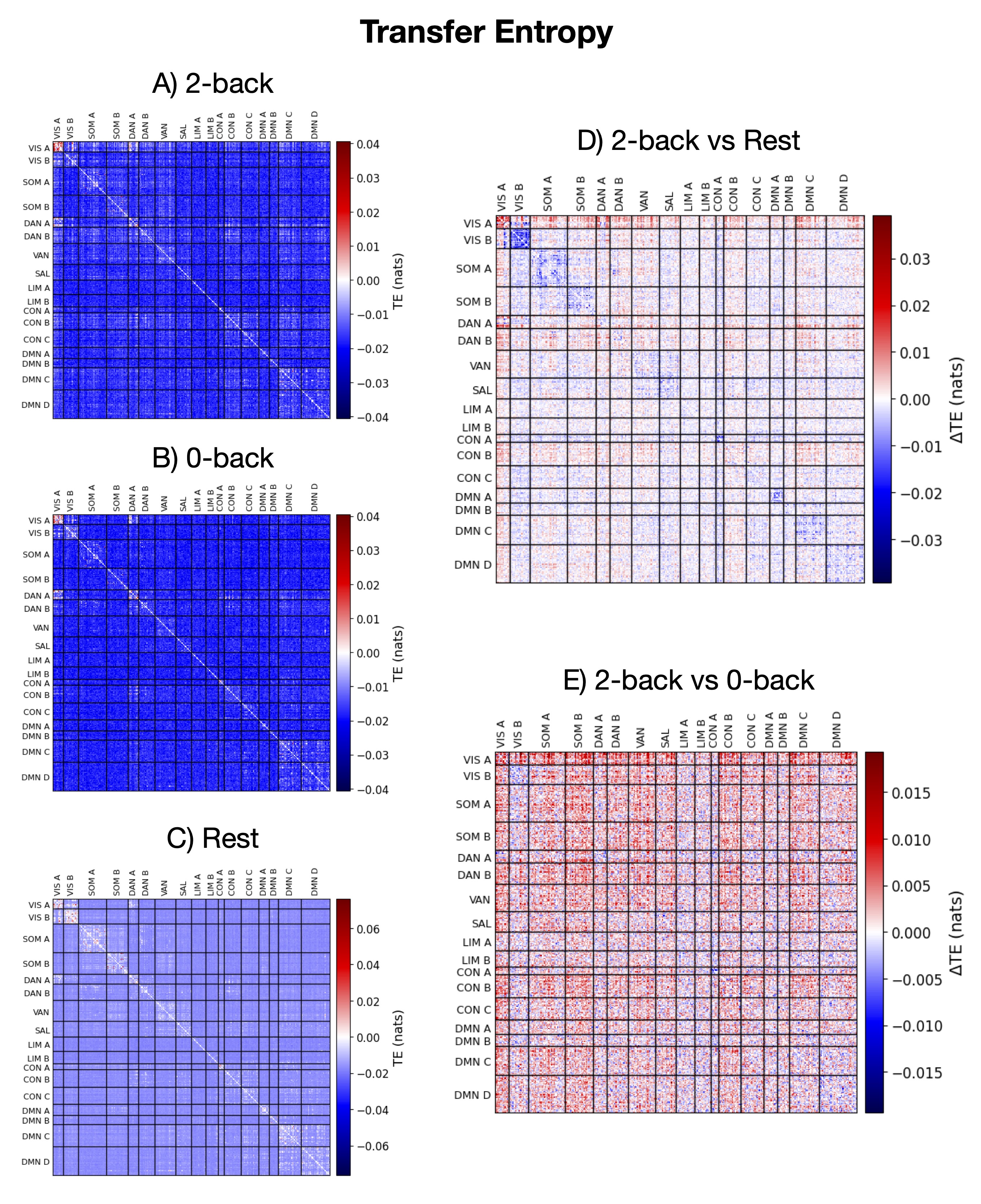}
    \caption{\textbf{Parcel-level transfer entropy matrices for each condition and contrast}.}
    \label{fig:all-te}
\end{figure}

\begin{figure}[hbt!]
    \centering
    \includegraphics[width=\textwidth]{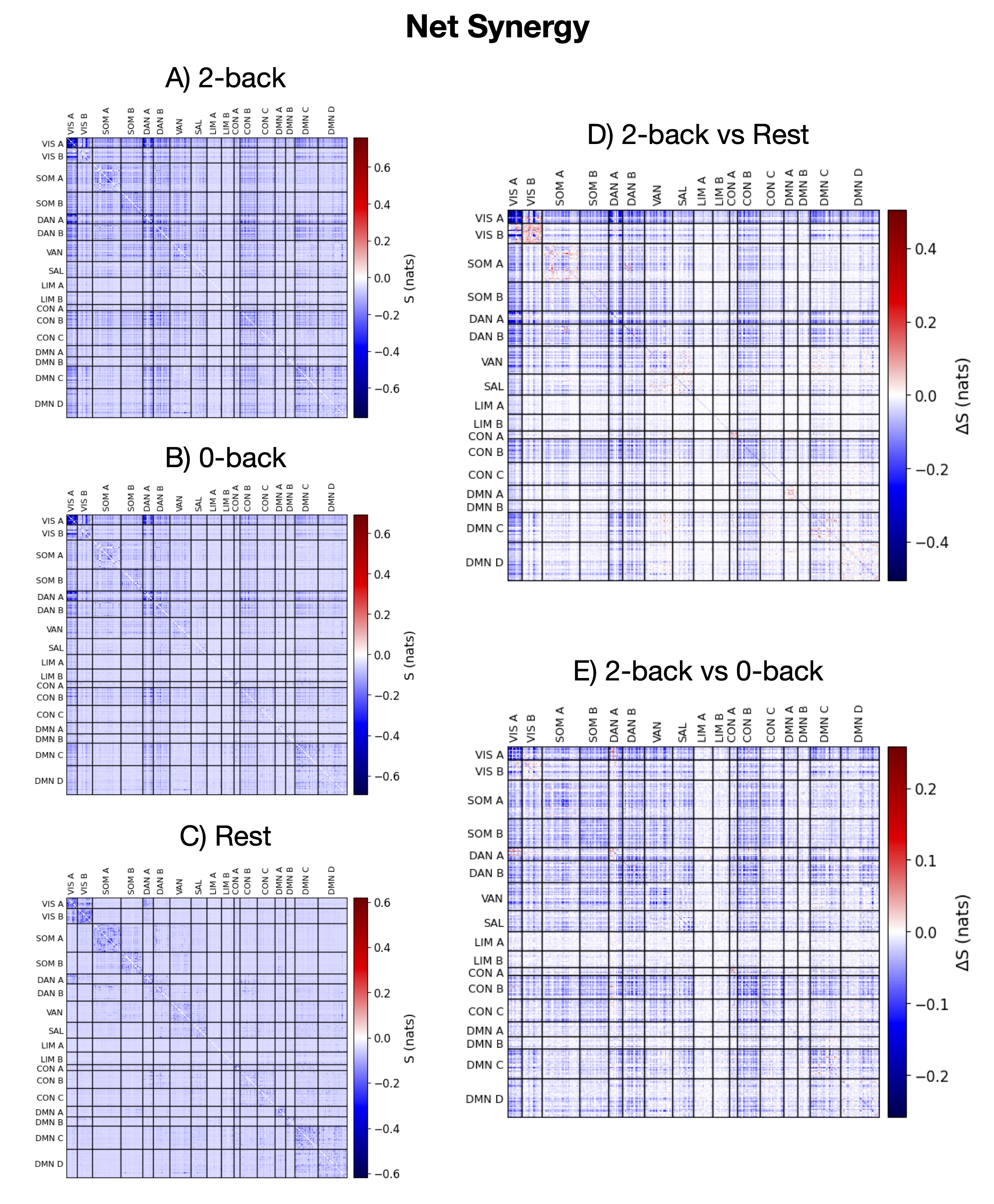}
    \caption{\textbf{Parcel-level net synergy matrices for each condition and contrast}.}
    \label{fig:all-synergy}
\end{figure}

\end{document}